# Analyzing user behavior of the micro-blogging website SinaWeibo during hot social events[*]

Wanqiu Guan[1]　Haoyu Gao[2]　Mingmin Yang[2]　Yuan Li[1]
Haixin Ma[3]　Weining Qian[3]　Zhigang Cao[2]　Xiaoguang Yang[2]

(1 College of Engineering and Information Technology, University of Chinese Academy of Sciences, 100049, Beijing, China)
(2 Academy of Mathematics and Systems Science, Chinese Academy of Sciences, 100190, Beijing, China)
(3 Institute of Massive Computing, East China Normal University, 200062, Shanghai, China)
(Emails: guanwanqiu10@mails.ucas.ac.cn, gaosteveneric@gmail.com, yangmingmin2005@126.com,
liyuan110@mails.ucas.ac.cn, 51111500010@ecnu.cn, wnqian@sei.ecnu.edu.cn, zhigangcao@amss.ac.cn, xgyang@iss.ac.cn)

## Abstract

The spread and resonance of users' opinions on SinaWeibo, the most popular micro-blogging website in China, are tremendously influential, having significantly affected the processes of many real-world hot social events. We select 21 hot events that were widely discussed on SinaWeibo in 2011, and do some statistical analyses. Our main findings are that (i) male users are more likely to be involved, (ii) messages that contain pictures and those posted by verified users are more likely to be reposted, while those with URLs are less likely, (iii) gender factor, for most events, presents no significant difference in reposting likelihood.

## Introduction

Social Networking Service, growing at an astonishing rate in recent years, is an online service that inspires users to share their interests and activities, and hence build new or closer social relationships. Twitter and Facebook are the most popular worldwide Social Networking Service websites, while in China, SinaWeibo [1], a hybrid of Twitter and Facebook, is the most prevailing one. Until the middle of 2012, the number of registered users on SinaWeibo had reached 368 million [2], and the current volume of weibos (microblogging messages in SinaWeibo, like tweets in Twitter), is over 100 million per day [3]. People can get information, update statuses, share views, and communicate with other users on SinaWeibo.

SinaWeibo is very similar to Twitter in several aspects. It also has a length limit of 140 characters. Users can follow others, repost weibos to their homepages, and broadcast their own weibos to followers. Additionally, users are also allowed to insert graphical emoticons and attach URLs, pictures, music, video files in every post.  If users want someone to read their messages, they can use the format "@username" to give a reminding. There are "VIP" accounts on SinaWeibo, which are called "verified users" (this mechanism was followed by Twitter). Majority of the verified users are "elite" people in the real world. Many of them are public celebrities, and some others are organization accounts. Users on SinaWeibo who want to be verified should send an application and give a proof of their identities. If they get the official approval, there will be a yellow sign of "V" after their names. Nevertheless, there are also some differences between Twitter and SinaWeibo. For instance, users on SinaWeibo can comment under a message, and choose whether to broadcast to their followers. But on Twitter, users cannot comment without reposting. In cooperation with internet censorship of the Chinese Government, SinaWeibo has much stricter controls over the posts [4].

---

[*] The research is supported by the 973 Program (2010CB731405) and NNSF of China (71101140).

The spread and resonance of the views of users on SinaWeibo can generate tremendous power and even affect the process of some social events. For one instance, in 2009, Deng Yujiao [5], a 21-year-old pedicure worker, stabbed a local township director to death when she tried to rebuff the latter's sexual harassment. This case aroused public indignation for the official and sympathy for Deng Yujiao. Millions of netizens posted messages or went to the local town to support her, which contributed to freeing her without criminal penalties. For another instance, in August 2010, Li Mengmeng [6], who missed the application of College Entrance Examination because of a local staff's dereliction of duty, posted her story on SinaWeibo to seek for help. Thousands of netizens reposted her weibo, helping her to gain the College-admission letter at last. Most of our cases in this paper, such as Xiao Yueyue and Guo Meimei, are substantially influenced by public opinions on SinaWeibo.

We select 21 hot events, which are widely discussed on SinaWeibo in 2011, and empirically analyze their posting and reposting characteristics. In comparison to Twitter's hot topics, we find that the reposting ratio of event-related weibos on SinaWeibo is much higher. Males are more actively involved in these hot events than females. Then for each event, we divide related weibos into original ones and reposting ones, and analyze their characteristics for different factors, namely picture, URL, gender and verification status. It is found that, for each event, the proportion of verified users in original weibos is significantly higher than that in reposting ones. So is the proportion of male users. It is interesting to notice that in most of the 21 events, picture containing original weibos are more likely to be reposted, while those with URLs are less likely. Another finding is that original weibos posted by verified users are more easily to be reposted, while the gender factor has little effect on reposting likelihood. It is also shown that the distribution of reposted times fit to the power law and the reposting depths are exponentially distributed.

## Related Work

With the emergence of Twitter (2006), the studies of microblogging websites are attracting more and more researchers in recent years. For instance, Kwak et al. (2010) studied the topological features of Twitter, and indicated that Twitter serves more of a news media than a social network [7]. Bakshy et al. (2011) investigated the attributes and relative influences of 1.6M Twitter users, and found that users that either have been influential in the past or have a large number of followers are more likely to generate the largest information cascades [8]. Wu et al. (2011) tried to answer "who says what to whom" on Twitter. They classified elite users into celebrities, bloggers, representatives of media outlets and other formal organizations, and found significant homophily within these categories and different attention distributions on news topics [9]. Perra et al. (2012) constructed an activity-driven model to describe the structure features of dynamic network and took Twitter as one of their empirical data sources [10].

As China's largest microblogging website, SinaWeibo, born in 2009, also attracts much attention from researchers. In comparison of Twitter and SinaWeibo, Gao et al. (2012) analyzed the textual features, topics and sentiment polarities of micro-posts for these two microblogging websites, revealed significant differences between them [11]. Yu et al. (2011) examined the key trending topics on SinaWeibo, compared them with the observations on Twitter. They found that, on SinaWeibo, trends emerge almost completely attributed to reposts of entertainment content such as jokes and images, while on Twitter, the trends are always due to current global events and news stories [12]. Guo et al. (2011) studied the topological characteristics of SinaWeibo, found that one-way relationship is the most common form between users, and the radius of SinaWeibo is very short [13]. Chen et al. (2012) compared the behaviors of verified users and unverified ones on SinaWeibo, and discussed the evolutions of social networks for these two groups of users [14]. Chen et al. (2012) found that SinaWeibo has a low reciprocity rate, and users are more likely to follow people of the same or higher social status [15]. Shen et al. (2010) divided spam weibos into news and advertisements, explored the principles of spam spreading by ROST Content Mining and network analysis [16]. Yang et al. trained an effective classifier to automatically detect the rumors from a mixed set of true information and false information [17]. Zhang et al. (2012) proposed an approach to generate lurking user's profiles by its followees' activities and did extensive experiments to verify the effectiveness of the approach [18]. For other microblogging websites, Li et al. unveiled that topic preferences are very different between Tencent and Twitter users in both content and time consuming [19].

There is also much research about hot events on these microblogging websites. To improve the precision of events detection, Packer et al. (2012) investigated the usage of Twitter in discussing the "Rock am Ring Music Festival", and proposed an approach which uses structured semantic query expansion to detect the related event [20]. Song et al. (2010) built a novel framework that mines the associations among topic trends in Twitter by considering both temporal and location information [21]. Shamma et al. (2010) analyzed tweets sampled during the 2009 inauguration of Barack Obama on Twitter. They found that messages which are widely spread can be served as broadcasted announcements, and the different levels of conversation reflect the variations in the level of interest in the media event [22]. Sakaki et al. (2010) built an earthquake detection model using real-time data of earthquakes in Japan. Applying this model, their detection is much faster than the announcements released by the Japan Meteorological Agency (JMA) [23]. Qu et al. (2011) conducted a case study of SinaWeibo, investigating how Chinese netizens used SinaWeibo in response to Yushu Earthquake [24].

Our work also falls into the booming field of computational social science, which investigates social phenomena through the medium of computing and related advanced information processing technologies [25]. The availability of data and advanced tools in theory and modeling of complex networks provide an integrated framework to study the predictive power of techno-social systems [31]. Recently, there are more and more findings that focus on quantifying the collective human characteristics via dealing with the online behavior data, say [27-29].

## Materials and Methods

Through developing a distributed crawler, we crawled the data set via APIs provided by SinaWeibo. Firstly, we chose 32 prestigious users (11 public intellectuals and 21 academic scholars), obtaining their profiles, posted weibos, followers and followees. All the 32 users are quite influential, and rather active in discussing public affairs. Then for each of the followers and followees, we obtain the same set of information as above. This process was repeated three times, i.e. there are four layers of users. Note that due to the API restriction of SinaWeibo, for each user, at most 5000 followers are crawled, although for some users, especially the 32 seeds, their total numbers of followers (which are known to us) may be much larger than this number. And for those whose total number of followers exceeds 5000, their top 5000 followers (ranked by number of followers) are crawled. A total of 1.6 million users and 650 million original and reposted weibos (created during August 2009 and January 2012) are crawled.

To study hot events discussed in SinaWeibo, we use the 20 hot events selected by 2011 China Internet Public Opinion Analysis Report. We find that one of the events, The Libya Civil War, is made up of two separate ones: The Libya Armed Conflict and The Death of Gaddafi. So we divide it into the two independent events, and finally get 21 hot events in total (Table 1). For each event, we derive the set of effective keywords with the assistance of three tools, search recommendation of Baidu (http://tuijian.baidu.com), search recommendation of SinaWeibo (s.weibo.com), and JinHua Keyword Tool (http://www.1n11.com). Note that a keyword is effective only if it is recommended by all of the three tools. By doing this, we can ensure that all the keywords we use are most relevant to the events we study, avoiding unnecessary noises (meanwhile, admittedly, this may make our dataset not so complete). We get 83 keywords in total and filter weibos containing these keywords in 60 continuous days, which start from the seventh day before the outbreak of each event. At last we get a set of 5036499 weibos. The hottest event has more than 746 thousand weibos, and the least hot one has only 7517 weibos.

**Table 1. The 21 hot events.**

| Title | Acronym | Brief introduction |
|---|---|---|
| **Guo Meimei**[1] | **Guo** | The Red Cross Society of China caught up in corruption scandal when Guo Meimei, a 20-year old girl, showed her luxurious life on SinaWeibo and claimed to be the general manager of a company called Red Cross Commerce. |
| **7·23 Wenzhou Train Collision**[2] | **723** | Two high-speed trains collided each other on a viaduct in the suburbs of Wenzhou, Zhejiang province. 40 people killed, at least 192 injured, including 12 severe injuries. |
| **Yao Jiaxin Murder Case**[3] | **Yao** | Yao Jiaxin, an undergraduate in Xi'an, stabbed the victim Zhang Miao to death after hitting her in a traffic crash. Many people concern about this case for Yao's family background and whether his death penalty should be abolished. |
| **Libyan Civil War**[4] | **Libya** | An armed conflict in the North African state of Libya, fought between forces loyal to Colonel Muammar Gaddafi and those seeking to oust his government. |
| **2011 Japanese Earthquake**[5] | **JE** | A magnitude 9.03 undersea earthquake of the coast of Japan, at 14:46 JST on 11 March 2011. The most powerful one ever in Japan, causing 15878 deaths, 6126 injured and 2713 people missing. |
| **House Purchase Restriction**[6] | **HPR** | The policy issued by local governments in China's major cities, including Beijing, Shanghai and Guangzhou, to curb the overheating real estate market by imposing restrictions on purchasing qualification. |
| **Gaddafi's Death**[7] | **Gaddafi** | Muammar Gaddafi, the deposed leader of Libya, was killed on 20 October 2011, after captured by National Transitional Council forces in a culvert west of Sirte. |

---

[1] http://www.chinadaily.com.cn/opinion/2011-07/15/content_12912148.htm
[2] http://en.wikipedia.org/wiki/Wenzhou_train_collision
[3] http://en.wikipedia.org/wiki/Yao_Jiaxin_murder_case
[4] http://en.wikipedia.org/wiki/Libyan_civil_war
[5] http://en.wikipedia.org/wiki/2011_T%C5%8Dhoku_earthquake_and_tsunami
[6] http://wiki.china.org.cn/wiki/index.php/Purchase_restriction_order
[7] http://en.wikipedia.org/wiki/Gaddafi%27s_drain

| | | |
|---|---|---|
| **Qian Yunhui's Death**[8] | Qian | Qian Yunhui, a village head in Yueqing, Zhejiang province, was crushed to death by a front wheel of a truck nearby his village. The public suspected that he was deliberately killed for his long history of appealing against the local government, while the government announced that it was but a normal car accident. |
| **Xiao Yueyue Incident**[9] | Yue | Wang Yue, a three year old girl, was run over by two cars in a small road in Foshan, Guangdong Province. None of the 18 passersby gave any help-hand. |
| **Brown Meat Essence**[10] | BME | Shuanghui Group, one of the largest meat processing companies in China, was accused for their meat products containing Brown Meat Essence. |
| **Salt Panic**[11] | Salt | Rumors said that the nuclear accidents caused by Japanese earthquake contaminated the sea salt, and eating iodized salt can prevent people from radiation. This rumor led to the salt panic in several big cities of China, where people rushed to supermarkets buying much more than needed salt. |
| **Steve Jobs's Death**[12] | Jobs | Steve Jobs, the co-founder, chairman, and CEO of Apple Inc, died of respiratory arrest related to his metastatic tumor on October 5, 2011. A great many Chinese express their condolences online. |
| **Free Lunch Program**[13] | Lunch | Deng Fei, a journalist of PhoenixTV in China, launched a charity program on SinaWeibo, which aims to provide kids in rural areas with a free lunch. More than 26 million students have benefited from this program by the end of September 2012. |
| **Coruption of Liu Zhijun**[14] | Liu | Liu Zhijun, the former Railway Minister of China, was dismissed because of abusing his position power to receive a large amount of bribery. |
| **2011 Summer Universiade**[15] | 2011 SU | The 2011 Summer Universiade, the XXVI Summer Universiade, was hosted in Shenzhen, Guangdong. |
| **Dyed Steamed Bun Scandal**[16] | DSBS | The China Central Television (CCTV) reported that some supermarkets in Shanghai sell "Dyed Steamed Buns", which are added with coloring and preservatives. |
| **Launch of ShenZhou-8**[17] | Shen8 | Shenzhou 8 was an unmanned flight of China's Shenzhou program, launched successfully on November 1, 2012. |
| **Shanghai Subway Crash**[18] | SSC | Two Shanghai subway trains slammed into another, injuring more than 270 passengers. Twelve officials punished. |
| **Microblogging to Combat Child Trafficking**[19] | MCCT | SinaWeibo has been used to hunt for child kidnap victims, on which netizens upload the photos of young beggars and call for people to repost and to recognize and rescue abducted children. The Seven-year-old Peng Wenle is a successful case who was kidnapped three years ago and was rescued by SinaWeibo. |

---

[8] http://en.wikipedia.org/wiki/Qian_Yunhui
[9] http://en.wikipedia.org/wiki/Death_of_Wang_Yue
[10] http://news.bbc.co.uk/sport2/hi/judo/8672794.stm
[11] http://www.chinasmack.com/2011/stories/salt-panic-chinese-fearing-japan-radiation-rush-to-buy-salt.html
[12] http://en.wikipedia.org/wiki/Steve_Jobs
[13] http://www.chinadaily.com.cn/video/2011-09-29/content_13813120.htm
[14] http://en.wikipedia.org/wiki/Liu_Zhijun
[15] http://en.wikipedia.org/wiki/2011_Summer_Universiade
[16] http://english.cri.cn/6909/2011/04/29/1461s634960.htm
[17] http://en.wikipedia.org/wiki/Shenzhou_8
[18] http://www.nytimes.com/2011/09/28/world/asia/shanghai-subway-accident-injures-hundreds.html
[19] http://www.china.org.cn/video/2011-02/12/content_21905807.htm

| Forbidden City Theft[20] | FCT | Seven art pieces made of gold and jewels were stolen from Forbidden City. It was the first theft for the Palace museum in 20 years, which caused a fueled discussion by Chinese netizens. |
| Li's Major in French Open[21] | Li | On 4 Jun 2011, Chinese girl Li Na beat Francesca Schiavone in the French Open Final, won her first major and became the first Grand Slam Singles Champion born in Asia. |

## Results

For each event, we group all the related weibos into OW (original weibos) and RW (reposting weibos), and then divide OW into two subclasses: OR (OW that have been reposted at least once) and ONR (OW that have never been reposted). It can be observed from Figure 1 that 13 out of all the 21 events are discussed in more than 100 thousand weibos. Taking into account that there may be quite a few replies behind each weibo, and much more users prefer reading but not to reply or repost, this amount shows that most of these events are indeed hot.

As we can see in Figure 1, the reposting ratios are all around 70.0% and the average ratio is 65.9%. This is more than twice of the average retweeting ratio of hot topics on Twitter, which is less than 31% (Haewoon Kwak, 2010 [7]). Thus compared with Twitter users, Chinese netizens prefer to repost weibos rather than creating original ones when talking about social events.

We can also see that in most events, the ratios of OR weibos are higher than 30.0%, and the average is 35.9%. This is a quite high level, because it is known that on Twitter there are only 6.0% tweets that have been retweeted [30].

Original weibos in our sample are the information sources of hot events, while the reposting ones represent the diffusion process. On the content dimension, picture and URL are the two basic factors, while on the identity dimension, gender and verification status are the elementary ingredients. In the following discussions, we shall focus on the above four factors.

---

[20] http://www.bbc.co.uk/news/world-asia-pacific-13356725
[21] http://en.wikipedia.org/wiki/Li_Na_(tennis)

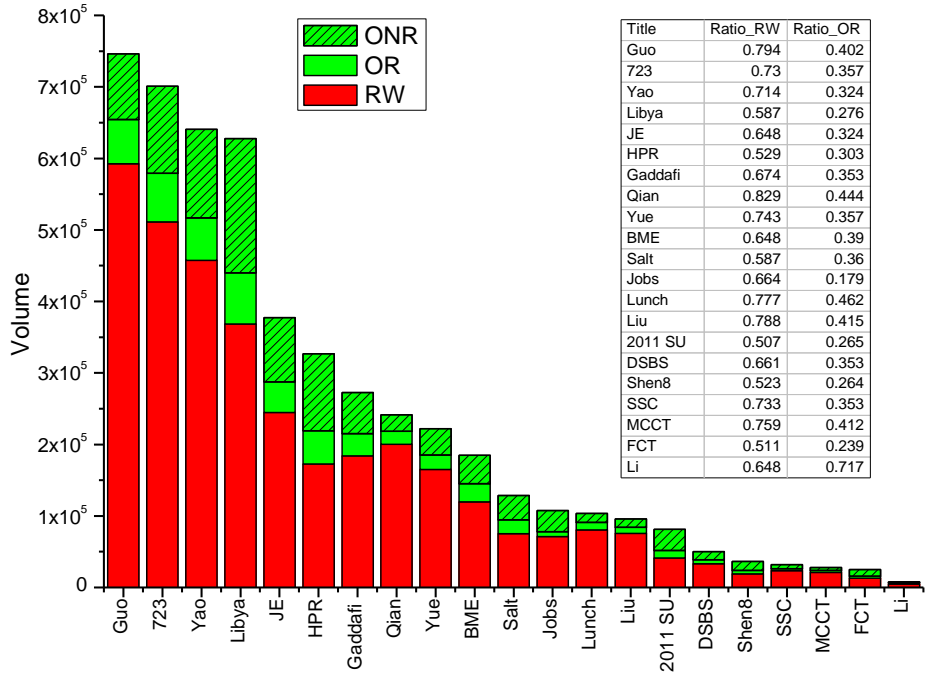

**Figure 1. Distribution of the amount of weibos for each event: RW shorts for reposting weibos, OR stands for original weibos that have been reposted at least once, and ONR represents those that have never been reposted. The detailed description of each event is presented in Table 1.**

## Picture & URL

On SinaWeibo, it is not allowed to add pictures when reposting and very few users add URLs due to the cumbersome operation. So we simply discuss picture and URL effects on original weibos. Overall, as shown in Figure 2, picture proportion of the 21 events ranges from 10% to 40%, with an average of 26%. The URL proportions are between 20% and 40%, with an average of 30%, which is a little higher than that of pictures.

We sort the 21 events in decreasing order of their total amounts of related weibos, and find that each of the top 9 events has a higher URL proportion than picture proportion, while for 10 of the rest 12 events the opposite property can be observed (The events of Jobs and Liu are the two exceptions ). To see why this happens, we further observe that almost all URLs are linked to news reports in the portal websites, like sina.com, sohu.com, and 163.com (this reflects the media attribute of SinaWeibo). So a higher proportion of URL implies a higher degree of media involvement. This answers our question because nowadays portal media are in general much more influential than ordinary SinaWeibo users (who can be taken as self media). To strengthen the above finding, we calculate two Pearson correlation coefficients, one for the amount of all weibos and the number of URL-containing original weibos (0.882), and the other for the amount of all weibos and the number of picture-containing original weibos (0.755). Both coefficients are positive, and the former is remarkably larger than the latter, indicating that the number of URLs is more positively correlated with the hotness of events. The events of Jobs and Liu, as the only two exceptions, caused much public attention but had relatively short life-spans. So the two events show special features of a relatively low weibo amount and high URL proportion.

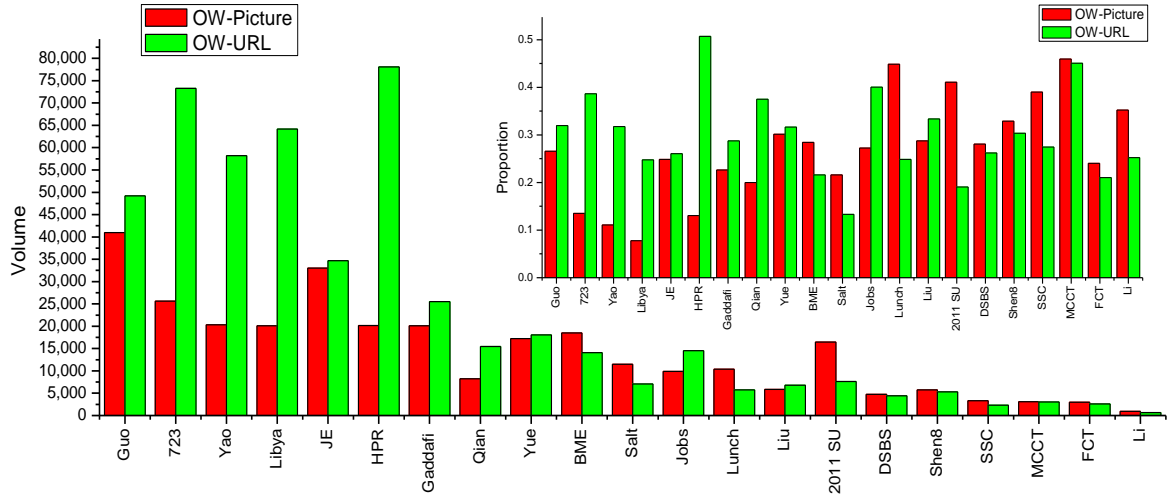

**Figure 2. Volumes and proportions of pictures and URLs.** In the upper embedded graph, OW-Picture refers to the picture proportion of original weibos, OW-URL denotes the URL proportion of original weibos. The detailed description of each event is presented in Table 1.

## Gender & verification status

For each event, we calculate the amount of original weibos posted by males and females, respectively, and reposting weibos by males and females, respectively. It is interesting to notice that the ratio of male users in our data is always larger than the overall ratio in SinaWeibo (57.4%, DCCI, 2011), with an average of 66.17%. We can conclude that, compared with females, Chinese males are more interested in hot events. And the proportion of males in original weibos is always higher than that in reposting weibos (Figure 3). What's more, in 20 out of all the 21 events (except for Salt), the proportion of males in original weibos is higher than the overall male ratio (66.17%), while in 17 out of all the 21 events, the male ratio of reposting weibos is lower than 66.17%. It indicates that in most of the 21 events, compared with females, male users prefer to create original weibos rather than reposting.

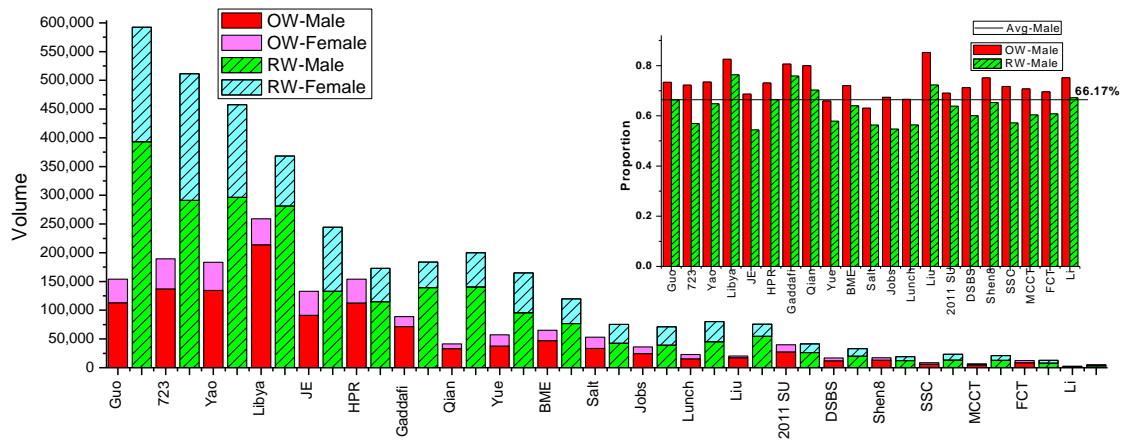

**Figure 3. Volumes and proportions of male users.** In the upper embedded graph, OW-Male represents the proportion of male users in the original weibos, RW-Male refers to that in reposting weibos. The definitions of RW-Female and OW-Female are similar. The detailed description of each event is presented in Table 1.

As in the gender analysis, we do the same treatment for verification status. We find that the proportion of verified users in all weibos of our sample is 23.5%, while for original ones, the proportion is nearly 30.0%, and for reposting ones, the proportion is only 17.5%. This is a strikingly high number because the overall ratio of verified users in SinaWeibo is only 0.1%. This suggests a very high participation ratio of verified users in hot events. For each event, we also calculate the verified proportions in original and reposting weibos respectively. As is shown in Figure 4, the verified proportions of original weibos are always higher than that of reposting ones in each event. We find that in 19 out of the 21 events (except for Yao and Jobs), the verification proportions of original weibos are higher than 23.5%, and also in 19 out of the 21 events (except for HPR and MCCT), the verification proportions of reposting weibos are lower than the baseline. We can say that in most events, compared with unverified ones, verified users prefer to post original weibos rather than reposting.

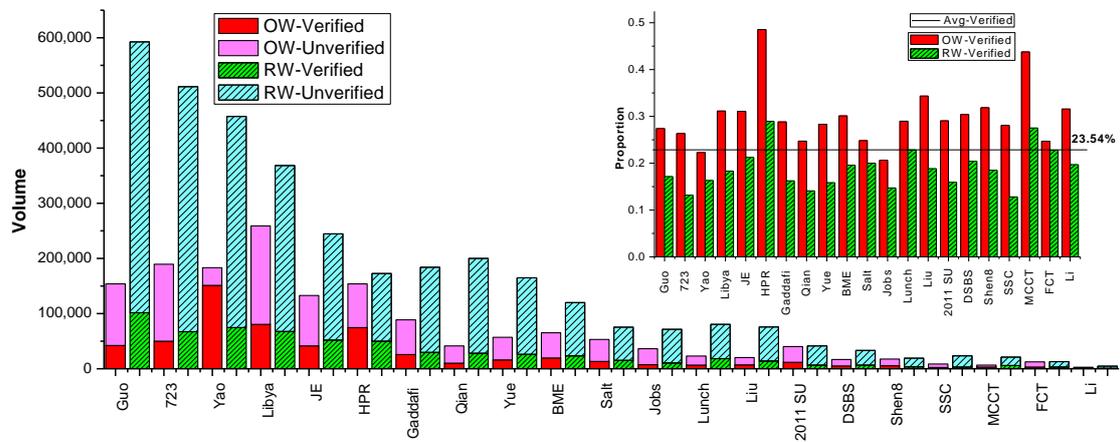

**Figure 4. Volumes and proportions of verified users.** In the upper embedded graph, OW-Verified /OW-Unverified represents the proportion of verified/ unverified users in the original weibos, RW-Verified/RW-Unverified refers to that in reposting weibos. The detailed description of each event is presented in Table 1.

To end this section, we make some comparisons between events in Table 2, and give some more empirical analyses (which require deeper understanding of the events).

**Table 2. The event comparisons in original weibos.**

|  | Top 3 events | Bottom 3 events |
|---|---|---|
| **Picture Proportion** | Combating Child-trafficking on SinaWeibo，Free Lunch，2011 Summer Universiade | Libya Civil War, Yao Jiaxin Murder Case, Guo Meimei |
| **URL Proportion** | House Purchase Restriction, Combating Child-trafficking on SinaWeibo, Steve Job's Death | Salt Panic, 2011 Summer Universiade, Brown Meat Essence |
| **Male Proportion** | Corruption of Liu Zhijun, Libya Civil War, Gaddafi's Death | Salt Panic, Xiao Yueyue Incident, Free Lunch |
| **Verified Proportion** | House Purchase Restriction, Combating Child-trafficking on SinaWeibo, Corruption of Liu Zhijun | Steve Job's Death, Yao Jiaxin Murder Case, Forbidden City Theft |

**1) Picture.** Each of the three events with the highest proportion of pictures consists of a series of smaller events and has various information sources. For example, MCCT is made up of a large number of child trafficking combating events, each of which may contain one or more pictures. This is also true for Lunch and 2011 SU. The three events with the lowest proportion of pictures, in contrast, are all single subjects.

**2) URL.** Compared with the bottom ones, the top three events have stronger news features, justifying our previous view that URLs are closely related to portal media.

**3) Gender.** From this angle, we can clearly find the differences between men and women. The top 3 events are all social and political problems, while the bottom 3 are related to daily lives (Salt Panic) or compassion (Xiao Yueyue and Free Lunch).

**4) Verification Status.** We believe that verified users are able to represent China's elites to some extent. Most of them live in metropolises (say Beijing and Shanghai), and thus are fond of discussing the House Purchase Restriction, which greatly influences their livings. They repost much more messages about "Combating Child-trafficking on SinaWeibo" than ordinary users, because in general they have much more followers and the victim parents prefer to "@" them for help. They are relatively wealthy, mostly in support of further political reforms, and thus more concerned with government corruptions.

## Reposted proportion & reposted times

In our dataset, some of the original weibos are reposted while the others are not. Some are reposted by quite a lot of people (with a maximum of about 37,894) while some others very few.

From Figure 5, we can see that the reposted proportions of picture-containing original weibos range from 30% to 60%, with an average of 53.93%, which is of a very high level. Those of URL-containing original weibos are much lower, ranging from 10% and 44%, with an average of 29.31%. In each event, the picture-containing original weibos possess a higher reposted proportion than the average, while the URL-containing original weibos have apparently a lower reposted proportion. This indicates that pictures are in favor of being reposted, and the effect of URLs is on the negative side.

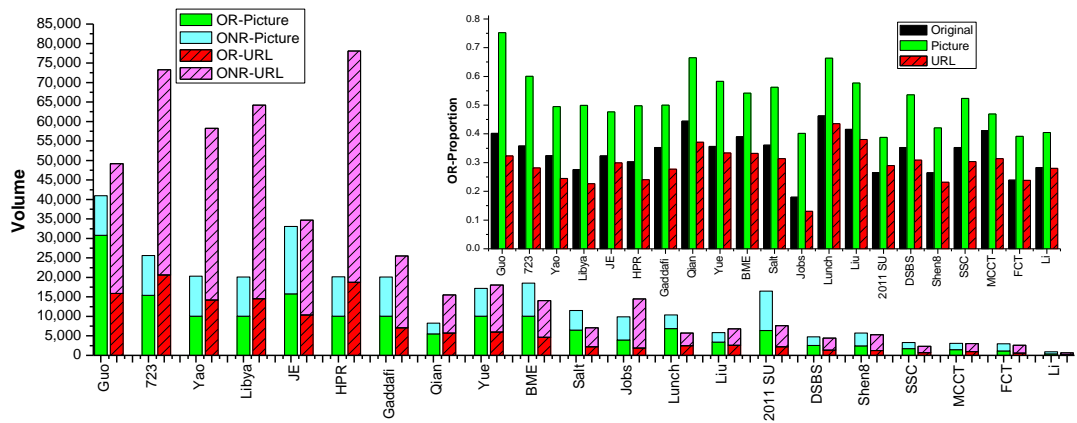

**Figure 5. Volumes and reposted proportions of URL/picture-containing original weibos.** OR-URL stands for the URL-containing weibos of OR (original weibos that are reposted), ONR-URL stands for the URL-containing weibos of ONR (original weibos that are not reposted). The definitions of OR-Picture and ONR-Picture are similar. And the upper embedded graph shows the reposted proportions. The detailed description of each event is presented in Table 1.

Similarly, we find that weibos posted by verified users are more easily to be reposted. Gender, however, has no effect at all for each event.

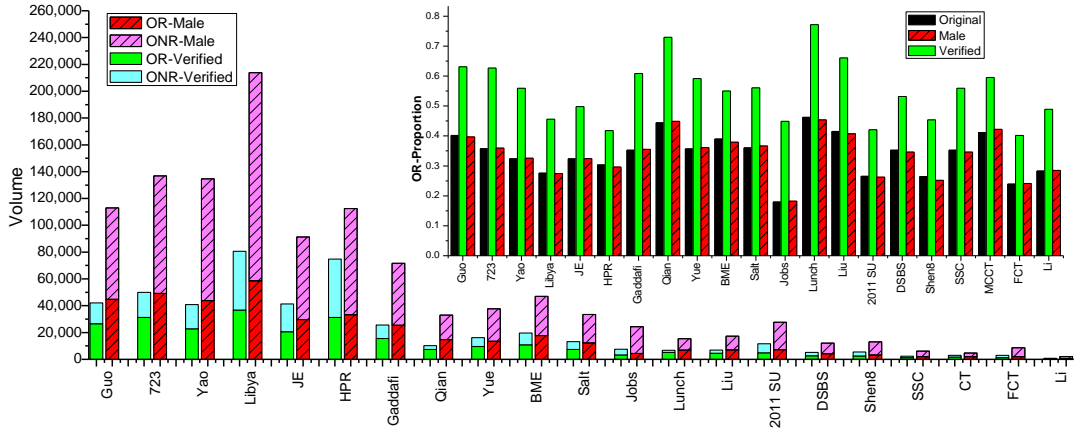

**Figure 6. Volumes and reposted proportions of picture/URL-containing original weibos.** The upper embedded graph shows the reposted proportion. OR-male stands for OR (original weibos that are reposted) that are created by male, and ONR-male stands for ONR (original weibos that are not reposted) that are created by male. OR-Verified and ONR-Verified are similarly defined. The detailed description of each event is presented in Table 1.

We also calculate the average reposted times for each factor, as shown in Figure 7. It can be easily seen that in each event, the reposted times of picture-containing original weibos are much larger than that without pictures, which verifies that picture is, in general, beneficial for reposting. URL-containing original weibos, however, have larger reposted times on average than non-URL ones. It indicates that although the reposted proportion is lower than average, once the URL-containing weibos is reposted, it would be forwarded more frequently than ordinary ones. The reposted times of weibos posted by verified users is significantly larger than those of unverified ones, while for weibos posted by males and females, respectively, their average reposted times are not far from each other. If weibos that are not reposted at all, whose reposted times are zero, are also taken into account in calculating the average reposted times, then the impact of URL is not clear.

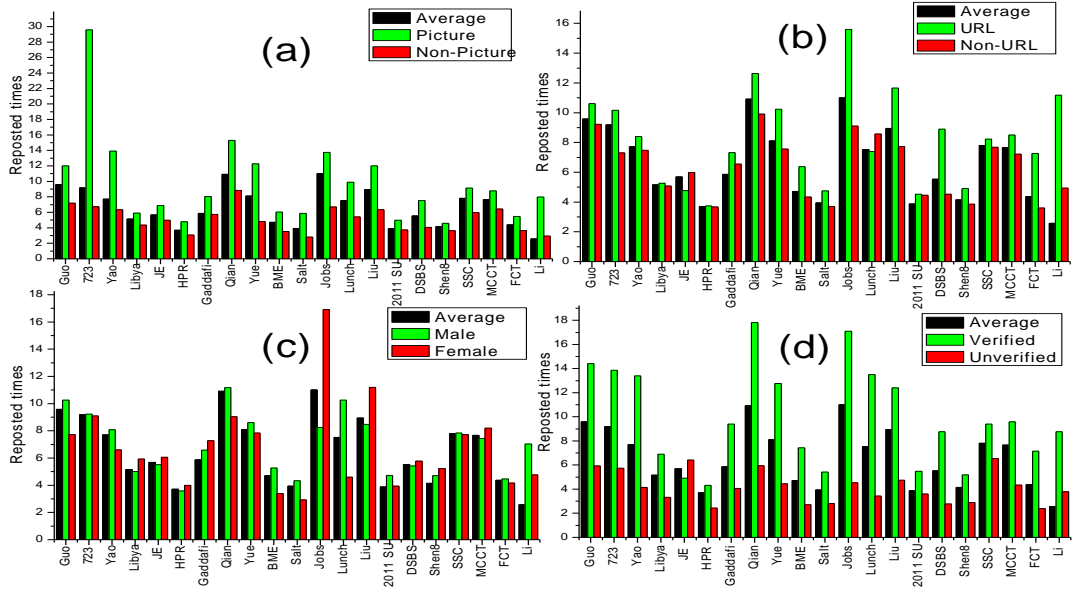

**Figure 7. Distributions of reposted times for each event classified by different factors**. (a): Picture, (b): URL, (c): Gender, (d): Verification status. The detailed description of each event is presented in Table 1.

In addition, for each event we calculate the correlation coefficients between reposted times and the following three factors: the total numbers of user's followers, followees and weibos. It is shown that the Spearman correlation coefficients are extremely low. This finding is different from previous conclusions [24][31-34].

**Table 3. Correlation coefficients between reposting times and three other variables.**

|  | **Guo** | **723** | **Yao** | **Libya** | **JE** | **HPR** | **Gaddafi** | **Qian** | **Yue** | **BME** | **Salt** |
|---|---|---|---|---|---|---|---|---|---|---|---|
| **Followers** | 0.043*** | 0.047*** | 0.034*** | 0.079*** | 0.083*** | 0.117*** | 0.087*** | 0.115*** | 0.174*** | 0.063*** | 0.041*** |
| **Followees** | 0.013*** | 0.007 | 0.005 | 0.009** | 0.023*** | -0.008 | -0.003 | 0.003 | 0.013** | 0.006 | 0.024*** |
| **Weibos** | 0.026*** | 0.042*** | 0.024*** | 0.041*** | 0.070*** | 0.079*** | 0.034*** | 0.048*** | 0.040*** | 0.053*** | 0.046*** |

|  | **Jobs** | **Lunch** | **Liu** | **2011 SU** | **DSBS** | **Shen8** | **SSC** | **MCCT** | **FCT** | **Li** |
|---|---|---|---|---|---|---|---|---|---|---|
| **Followers** | 0.144*** | 0.010*** | 0.057*** | 0.073*** | 0.127*** | 0.102*** | 0.125*** | 0.131*** | 0.161*** | 0.128*** |
| **Followees** | 0.030*** | -0.002 | 0.014*** | 0.031*** | 0.007 | 0.008 | 0.007 | 0.079*** | 0.034*** | 0.101*** |
| **Weibos** | 0.180*** | 0.005*** | 0.054*** | 0.056*** | 0.106*** | 0.059*** | 0.060*** | 0.114*** | 0.141*** | 0.290*** |

\* Represents statistical significance at the 0.10 level (2-tailed).

\*\* Represents statistical significance at the 0.05 level (2-tailed).

\*\*\* Represents statistical significance at the 0.01 level (2-tailed).

From the scatter plot of Figure 8, we can find that there is no obvious relation between reposted times and number of followers. Nevertheless, if we only consider the weibos posted by users with 500,000 followers or less and the corresponding maximum reposted times, then positive linear correlation does exist.

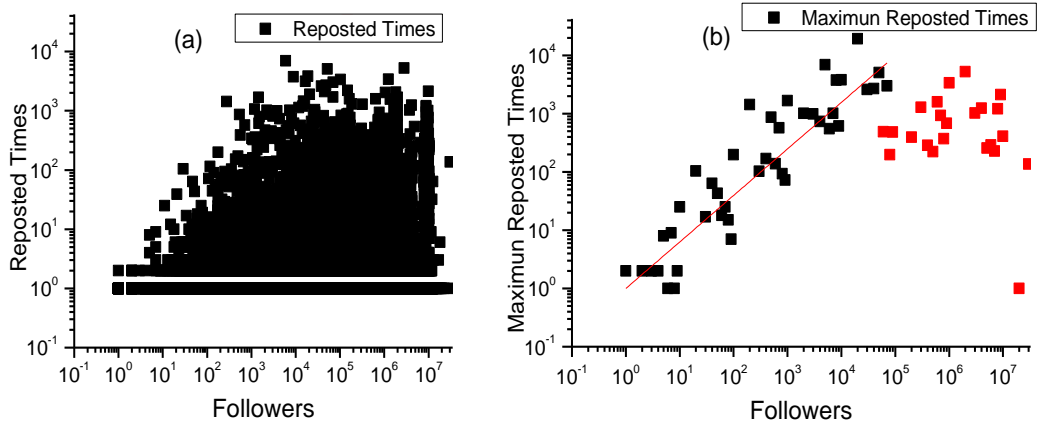

**Figure 8. Relations between number of followers and reposted times**. (a): scatter plot between the reposted times and number of followers, (b): the positive linear relation between the corresponding maximum reposted times and number of followers.

To validate the above findings, we present the OLS regression results for each event. Firstly, we take a log transformation on reposted times and use it as the dependent variable, and choose the seven factors discussed above (three normal variables, i.e. the numbers of users' followers, followees, weibos, and four 0-1 variables, i.e. gender, the verification status, picture-containing, and URL-containing) as independent variables. Note that we take "male" as value 1 and "female" as 0. As shown in Table 4, in all the 21 events, verification, picture and URL all show significant effect. The coefficients of verification and picture are positive, indicating that these two variables are in favor of being reposted. However, the coefficient of URL presents significantly negative, implying that weibos containing URLs are unlikely to be reposted. For gender, we can see that in 11 of the 21 events, the coefficients are insignificant, indicating that the gender factor has little impact on the likelihood of being reposted. For other control variables, most of them are significant and positively related with the probability of being reposted, which is in accordance with Table 3. In a word, our previous findings are further verified by the regressions.

**Table 4. The OLS regression of reposted times.**

| Variable | Guo | 7-23 | Yao | Libya | JE | HPR | Gaddafi |
|---|---|---|---|---|---|---|---|
| Intercept | 0.20*** | 0.13*** | 0.16*** | 0.11*** | 0.17*** | 0.18*** | 0.14*** |
| Followers | 9.70E-08*** | 1.91E-07*** | 1.55E-07*** | 1.34E-07*** | 2.01E-07*** | 2.09E-07*** | 1.42E-07*** |
| Friends | 8.29E-05*** | 4.59E-05*** | 8.04E-05*** | 5.55E-05*** | 1.20E-05** | 1.03E-05 | 6.69E-05*** |
| Status | 2.08E-05*** | 2.36E-05*** | 2.32E-05*** | 2.11E-05*** | 2.03E-05*** | 1.59E-05*** | 1.63E-05*** |
| Verification | 0.41*** | 0.35*** | 0.42*** | 0.29*** | 0.21*** | 0.22*** | 0.38*** |
| Gender | -0.01*** | -0.01*** | 0.01 | -0.01*** | -0.02*** | -0.02*** | -0.01 |
| Picture | 0.27*** | 0.26*** | 0.26*** | 0.16*** | 0.19*** | 0.19*** | 0.24*** |
| URL | -0.16*** | -0.11*** | -0.13*** | -0.13*** | -0.06*** | -0.20*** | -0.17*** |
| Observations | 153961 | 177486 | 183258 | 259035 | 174349 | 153965 | 88791 |
| Adjusted $R^2$ | 0.15 | 0.22 | 0.17 | 0.17 | 0.13 | 0.16 | 0.20 |

| Variable | Qian | Yue | BME | Salt | Jobs | Lunch | Liu |
|---|---|---|---|---|---|---|---|
| Intercept | 0.23*** | 0.12*** | 0.20*** | 0.17*** | 0.01** | 0.22** | 0.16*** |
| Followers | 1.27E-07*** | 1.71E-07*** | 1.22E-07*** | 3.65E-08*** | 2.92E-07*** | 6.66E-08*** | 7.81E-08*** |
| Friends | 2.07E-04*** | 5.42E-05*** | 4.80E-05*** | 4.96E-05*** | -5.05E-06 | 1.22E-04 | 1.11E-04*** |
| Status | 2.74E-05*** | 2.22E-05*** | 1.58E-05*** | 2.28E-05*** | 4.99E-05*** | 2.11E-05*** | 1.61E-05*** |
| Verification | 0.56*** | 0.34*** | 0.29*** | 0.29*** | 0.31*** | 0.45*** | 0.45*** |
| Gender | -0.003 | -0.02*** | -0.02*** | 0.01 | -0.05*** | -0.03 | -0.02 |
| Picture | 0.35*** | 0.38*** | 0.25*** | 0.26*** | 0.26*** | 0.30*** | 0.32*** |
| URL | -0.15*** | -0.10*** | -0.14*** | -0.13*** | -0.05*** | -0.19*** | -0.13*** |
| Observations | 41283 | 57058 | 65158 | 53135 | 36189 | 23092 | 20358 |
| Adjusted $R^2$ | 0.19 | 0.22 | 0.14 | 0.14 | 0.34 | 0.21 | 0.19 |

| Variable | 2011SU | DSBS | Shen8 | SSC | MCCT | FCT | Li |
|---|---|---|---|---|---|---|---|
| Intercept | 0.07*** | 0.17*** | 0.10*** | 0.12*** | 0.18*** | 0.10*** | 0.07*** |
| Followers | 1.99E-07*** | 1.81E-07*** | 1.58E-07*** | 2.06E-07*** | 2.32E-07*** | 1.98E-07*** | 1.02E-07*** |
| Friends | 5.43E-05*** | 5.46E-05*** | 2.20E-05*** | 6.37E-05*** | 1.41E-04*** | 5.44E-06*** | 3.54E-05*** |
| Status | 1.83E-05*** | 1.62E-05*** | 1.11E-05*** | 1.82E-05*** | 2.33E-05*** | 1.27E-05*** | 4.51E-05*** |
| Verification | 0.19*** | 0.32*** | 0.23*** | 0.28*** | 0.39*** | 0.24*** | 0.33*** |
| Gender | -0.01 | -0.03** | -0.02** | 0.01 | -0.02 | 0.01 | -0.01 |
| Picture | 0.18*** | 0.24*** | 0.19*** | 0.38*** | 0.19*** | 0.20*** | 0.14*** |
| URL | -0.03*** | -0.11*** | -0.12*** | -0.23*** | -0.22*** | -0.09*** | -0.07*** |
| Observations | 40064 | 16939 | 17441 | 8481 | 6730 | 12339 | 2646 |
| Adjusted $R^2$ | 0.13 | 0.16 | 0.21 | 0.21 | 0.19 | 0.16 | 0.19 |

\* Represents significance at the 0.10 level (2-tailed).

\*\* Represents significance at the 0.05 level (2-tailed).

\*\*\* Represents significance at the 0.01 level (2-tailed).

**Distributions of reposted weibos**

In this subsection we study the distributions of the reposted times and reposted depths. We find that in each event, reposted times fit to a power law distribution, as shown in Table 5. It indicates that most of the original weibos are reposted for a few times, while very few ones are reposted for a large amount, and the distribution is scare-free. For instance, in the event 723 Wenzhou Incident, there are 88.55% weibos that are reposted for at most once, including 64.26% that are never reposted, while there are only about 0.03% weibos that are reposted for more than 1,000 times.

We draw the distribution of 723 Wenzhou Incident as an example (Figure 9). We can find that for weibos that are reposted for less than 500 times, the number of URL containing ones is always smaller than that of picture. What's more, for those reposted for 500 times or more, we calculate that there are 68 picture containing ones, with a total reposted amount of 110141 , and only 31 URL containing ones, with a total reposted amount of 37183. In this case, we can conclude that original weibos with pictures are more easily to be reposted than those with URLs.

**Table 5. Coefficients in the power-law distributions of reposted times.**

|      | Guo    | 723    | Yao    | Libya  | JE     | HPR    | Gaddafi | Qian   | Yue    | BME    | Salt   |
|------|--------|--------|--------|--------|--------|--------|---------|--------|--------|--------|--------|
| beta | -0.996 | -1.291 | -0.494 | -0.772 | -0.935 | -1.131 | -1.153  | -1.109 | -0.821 | -0.912 | -1.021 |
| $R^2$ | 0.7161 | 0.8256 | 0.5428 | 0.7028 | 0.7046 | 0.7551 | 0.7946  | 0.7524 | 0.659  | 0.7547 | 0.7358 |

|      | Jobs   | Lunch  | Liu    | SUG    | DSBS   | Shen8  | SSC    | MCCT   | FCT    | Li     |
|------|--------|--------|--------|--------|--------|--------|--------|--------|--------|--------|
| beta | -0.753 | -0.962 | -0.875 | -0.824 | -1.127 | -1.167 | -1.114 | -0.932 | -0.926 | -0.942 |
| $R^2$ | 0.6139 | 6538   | 0.6836 | 0.6409 | 0.7209 | 0.7728 | 0.7394 | 0.6843 | 0.7297 | 0.7032 |

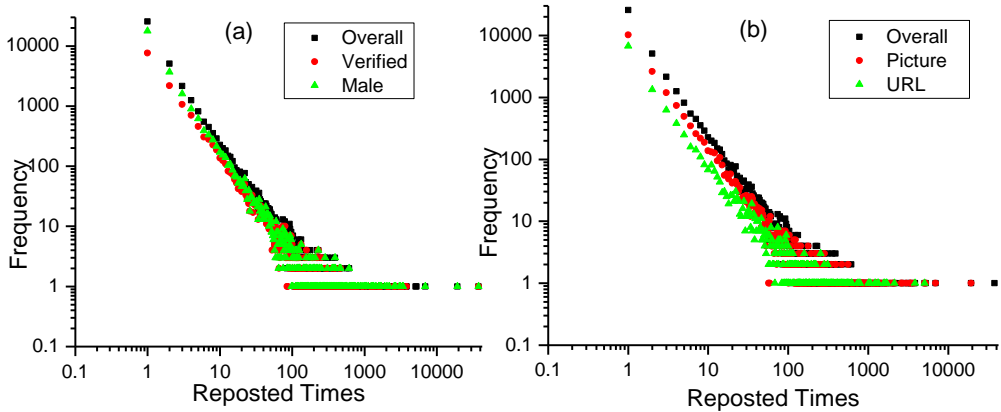

**Figure 9. The power law distribution of reposted times of** 723 Wenzhou Incident. (a): the distributions of reposted times for factors Male and Verification status, (b): the distributions of reposted times for factors Picture and URL.

The depths of reposting trees, however, are well fitted into exponential distributions as shown in Figure 10 (with average R squared 0.987).

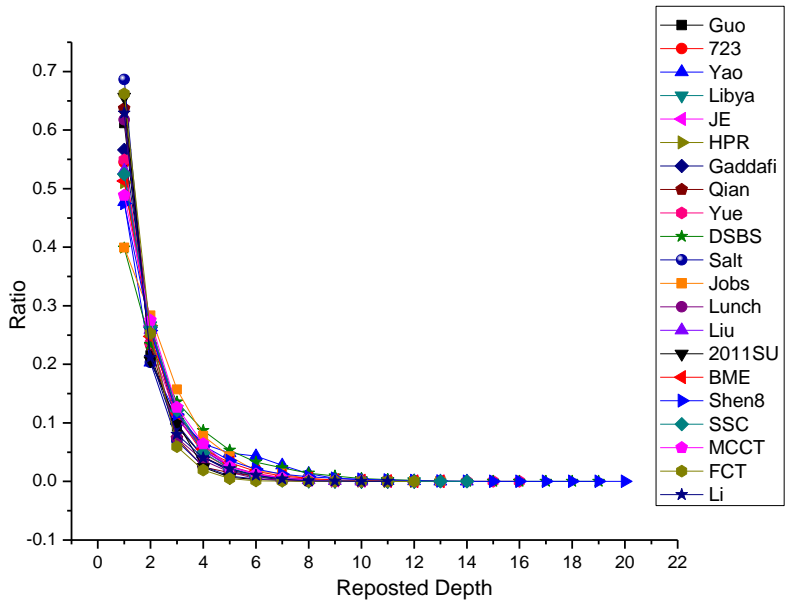

**Figure 10. The exponential distribution of reposted depths.** The detailed description of each event is presented in Table 1.

We select all the weibos with reposted depths 9 or larger (which are referred to as long-chain weibos, totally sum to 6911, just accounting for 0.5% of RWs), and do a manual text analysis. We observe that most of the long-chain weibos are written in the second day of the incident or later, and find that they can be roughly classified into five types.

**1) Asking for repost.** Weibos of this type contain phrases such as "Help to repost it!" or "If you support me, please repost this message!" These weibos are the most among all the five types (35.29%).

**2) Queuing repost.** When a user reposts the original weibo with a very incisive or special comment, other users will sequentially repost the message of the former and repeat this comment. For example, in the event of Xiao Yueyue, one netizen posted "Please end the cold-heartedness". Then thousands of people reposted from him and added the same words, forming a regular queue, as shown in Figure 11. This type accounts for 30.59% of the long-chain weibos.

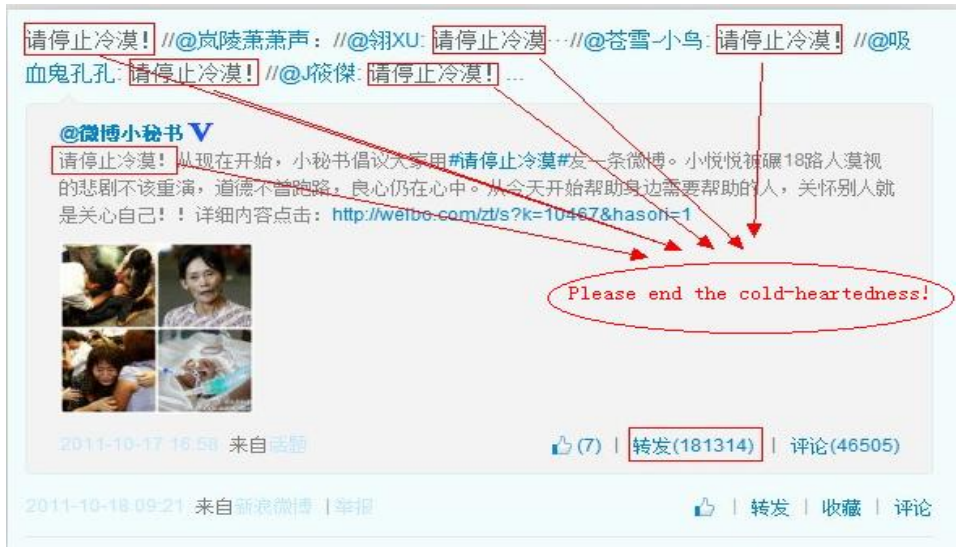

**Figure 11. An example of queuing repost on SinaWeibo.**

**3) New information/views.** When new information or new views occur, it is also likely to produce a deeply reposted weibo. But the amount of this type of weibos is much smaller than the former two, with a proportion of 17.65%.

**4) Humors.** Talented users create weibos to banter or satirize the current affairs (11.76%).

**5) Conversations.** Very seldom, people chat on SinaWeibo (4.71%).

## Discussion

In this paper we discussed the posting and reposting characteristics of 21 hot events on SinaWeibo. We found that the average reposting proportion of hot events in SinaWeibo is much higher than the tweeted proportion of hot topics in Twitter, and the average proportion of male users involved in hot events is much higher than the overall proportion of male users in SinaWeibo. For each of the 21 events, we concluded that: (i) The proportion of verified users in all the original weibos is significantly higher than that in reposted ones. So is the proportion of male users. (ii) Picture containing original weibos are more likely to be reposted, while those with URLs are less likely. (iii) Original weibos posted by verified users are more likely to be reposted, while the gender factor affects very little on reposted likelihood. We also studied the distribution characteristics of reposted weibos, and derived the power law distribution of reposted times and the exponential distribution of depths of reposted chains. Among the 21 events, three of them (namely Guo, Lunch and MCCT) are endogenously triggered, and all the others are exogenously triggered. We ran all the processing separately on them, and did not observe any significant difference. In the future, we shall focus more on the network characteristics and "tipping" mechanisms of the hot events.